\documentclass[final,1p,times,twocolumn]{elsarticle}

 \usepackage{graphicx}

\usepackage{amssymb}

\usepackage[colorlinks,dvipdfm]{hyperref}

\begin{document}

\begin{frontmatter}



\title{Analysis and improvement of a strongly secure certificateless key exchange protocol without pairing}


\author[label1,label2]{Min Zhang}
\ead{zhangmin@bupt.edu.cn}
\author[label1,label2]{Jie Zhang}

\author[label2]{Qiao-Yan Wen}
\author[label2]{Zheng-Ping Jin}
\author[label2]{Hua Zhang}

\renewcommand{\thefootnote}{\fnsymbol{footnote}}

\address[label1]{School of Science, Beijing University of Posts and Telecommunications, Beijing, 100876, China}
\address[label2]{State Key Laboratory of Networking and Switching Technology, Beijing University of Posts and Telecommunications,

Beijing, 100876, China}

\begin{abstract}

Recently, Yang and Tan proposed a certificateless key exchange protocol without pairing, and claimed their scheme satisfies forward secrecy, which means no adversary could derive an already-established session key unless the full user secret keys (including a private key and an ephemeral secret key) of both communication parties are compromised. However, in this paper, we point out their protocol is actually not secure as claimed by presenting an attack launched by an adversary who has learned the private key of one party and the ephemeral secret key of the other, but not the full user secret keys of both parties. Furthermore, to make up this flaw, we also provide an improved protocol in which the private key and the ephemeral secret key are closely  intertwined with each other for generating the session key, thus above attack can be efficiently resisted.

\end{abstract}


\end{frontmatter}


\section{Introduction}
\label{1}

In traditional public key cryptography (PKC), a trust certification authority (CA) signs a digital certificate of a user, and the public key infrastructure (PKI) manages the certificate to provide the authenticity of public keys. However, certificate management, including distribution, revocation, storage and validation cost, should face many challenges in practice [1]. To resolve the problem of certificate management, identity-based public key cryptography (ID-PKC) was proposed by Shamir [2] in 1984. Its basic idea is that the users can choose arbitrary strings, such as their email addresses or other
online identifies, as their public keys, and the corresponding private keys are created by binding the identities with a master key
of a trusted private key generator (PKG). In this case, there is no need for certification, but a new question came out. KGC is needed to make the private key for every user according to his identity, which means it can get all the users' secret keys. Thus, ID-PKC has to confront so-called key escrow problem. In order to eliminate the drawbacks of both ID-PKC and PKI, a new paradigm of certificateless public key cryptography (CL-PKC) was provided by Al-Riyami and Paterson [3] in 2003. The basic idea of CL-PKC is the construct of private key which is combining a partial private key generated by the KGC with some secret value chosen by himself. Obviously, the CL-PKC is more interesting as which received both benefits of the ID-PKC and traditional PKI. Thus, CL-PKC is often considered as a cross between PKI and ID-PKC.

\vspace{0.1cm}

 Key exchange (KE) protocols are mechanisms which establish a shared key by two or more parities communicating over an insecure network. However, compared with the certificateless encryption and signature [4-13], the study of key exchange protocol based on CL-PKC is seldom discussed. Al-Riyam and Paterson [3] proposed the first certificateless key exchange protocol which had no formal security model and proof. Later, some certificateless key exchange (CL-KE) protocols [1,14,15] were proposed with heuristic key security analysis. Then, Swanson [16] gave the general security analysis to the proposed certificateless key exchange protocols. However, all of certificateless key exchange protocols above are based on the bilinear pairings. Compared with the exponentiations, the computation of pairing is extremely expensive, so the certificateless key exchange protocol without pairing based on the CL-PKC were proposed by Geng and Hou [17,18]. Unfortunately, none of these protocols is secure [19]. Recently, Yang and Tan [20] proposed a new CL-KE protocol without pairing and claimed that their scheme is strongly secure to their security model.

\vspace{0.1cm}

In this paper, we point out that Yang and Tan's protocol is actually not secure as claimed by presenting an attack launched by an adversary who has learned the private key of one party and the ephemeral secret key of the other, but not the full user secret keys of both parties. That is, the adversary can make a RevealEphemeralKey(A,i) query to learn the ephemeral secret key $e_{A}$ of one communication party A and make a RevealSecretValue(B) query to learn the private key $S_{B}$ of the corresponding party B, and successfully calculates the session key, which means that the forward secrecy is not satisfied. Furthermore, to make up this flaw, we also provide an improved protocol in which the private key and the ephemeral secret key are closely  intertwined with each other for generating the session key. In other words, we add $z_{8}=g^{(e_{B}+z_{B})(e_{A}+S_{A}+z_{A})}$ and $z_{9}=g^{(S_{A}+e_{A})(S_{B}+e_{B})}$ into the generated session key, such that any adversary can calculate neither $z_{8}$ nor  $z_{9}$ even if he knows the values of $e_{A}$ and $S_{B}$. Thus, the session key can not be computed and the protocol what we improved can effective avoid the attacks mentioned above.

\vspace{0.1cm}

The rest of this paper is organized as follows: In section 2, we list the certificateless key exchange protocol and its security model. In section 3, we review Yang and Tan's strongly secure certificateless key exchange protocol without pairing. In section 4, we give our attacks on Yang and Tan's scheme as well as a possible improvement . We give some further security discussions in section 5.  Finally, we conclude the paper in section 6.

\section{Certificateless key exchange and its security model}
\label{2}

\subsection{Certificateless  key  exchange}
\label{2.1}

A CL-KE protocol is specified by the following probabilistic polynomial time algorithms:
\vspace{0.2cm}

{\bf Setup $(1^{k})$}. This algorithm takes a security parameter $k$ as input and returns the master secret key $msk$ and the master public key $mpk$.

\vspace{0.2cm}

{\bf ExtractIdBasedKey$(msk,ID)$}. This algorithm takes master key $msk$ and a user's identity $ID$ as input, and returns a partial private key $D_{ID}$ corresponding to the user.

\vspace{0.2cm}

{\bf SetSecretValue$(mpk,ID)$}. This algorithm takes the master public key $mpk$ and a user's identity $ID$ as input, and returns secret value $S_{ID}$ corresponding to the user.

\vspace{0.2cm}

{\bf SetPublicKey$(mpk,D_{ID},S_{ID})$}. This algorithm takes the master public key $mpk$, a user's the secret values $S_{ID}$ as input, and returns a public key $pk_{ID}$ corresponding to the user.

\vspace{0.2cm}

{\bf SetPrivateKey$(mpk,D_{ID},S_{ID})$}. This algorithm takes a master public key $mpk$, a user's partial private key $D_{ID}$ and a  secret value $S_{ID}$ as input, and returns a full private key $sk_{ID}$ corresponding to the user.

\vspace{0.3cm}

\subsection{Adversarial model}
\label{2.2}

\ In CL-KE protocol as defined in [1], the adversarial model is defined via a game between an adversary $\mathcal{A}$ and a game simulator S. At first, S runs the setup algorithm to generate $(mpk,msk)$ and returns $mpk$ to $\mathcal{A}$. Then $\mathcal{A}$ can deliver, drop, modify or inject messages for he can control all the network. Furthermore, $\mathcal{A}$ may ask a polynomial number of the following queries:

\vspace{0.2cm}

{\bf CreateUser $(ID)$}. By this query, the adversary $\mathcal{A}$ sets up a new user with identity $ID$. Upon receiving such a query, $S$ generates $D_{ID},S_{ID},pk_{ID}$ and $sk_{ID}$, returns $pk_{ID}$ to $\mathcal{A}$.

\vspace{0.2cm}

{\bf Send$(U,i,m)$}. By this query, the adversary $\mathcal{A}$ input the message $m$ to instance $\prod_{U}^{i}$. $\prod_{U}^{i}$ executes protocol and returns the output message $M_{out}$ to $\mathcal{A}$.

\vspace{0.2cm}

{\bf RevealMasterKey$(U)$}. This query allows $\mathcal{A}$ to obtain the $msk$.

\vspace{0.2cm}

{\bf RevealIDBasedKey$(U)$}. This query allows $\mathcal{A}$ to learn the $D_{U}$.

\vspace{0.2cm}

{\bf RevealSecretValue$(U)$}. This query allows $\mathcal{A}$ to obtain the $S_{U}$.

\vspace{0.2cm}

{\bf RevealSecretKey$(U)$}. This query allows $\mathcal{A}$ to learn the $sk_{U}$.

\vspace{0.2cm}

{\bf RevealEphemeralKey$(U,i)$}. This query allows $\mathcal{A}$ to obtain the ephemeral secret key generated by $\prod_{U}^{i}$.

\vspace{0.2cm}

{\bf RevealSessionKey$(U,i)$}. This query allows $\mathcal{A}$ to learn the session key $ssk_{U}^{i}$ if $\prod_{U}^{i}$ accepted; otherwise, $\bot$ is returned.

\vspace{0.2cm}

{\bf ReplacePublicKey$(U,(pk_{U})')$}. This query allows $\mathcal{A}$ to replace $U's$ public key with $pk_{U}= (pk_{U})'$. After this query, $S$ will use the new key pair as $U's$ $public\setminus private$ key pair.

\vspace{0.2cm}

{\bf Test$(U^{*},i^{*})$}. This query allows $\mathcal{A}$ to select a challenge instance $\prod_{U}^{i}$ that has accepted. Upon receiving this query, a random coin $b$ is flipped by $S$. If the coin $b=1$, then $S$ return $ssk_{U^{\ast}}^{i^{\ast}}$ to $\mathcal{A}$. Otherwise, a random session key is drawn from the session key space and returned to the adversary. This query is only made once by $\mathcal{A}$ during the game, and $\prod_{U^{\ast}}^{i^{\ast}}$ must have accepted the conversation, and is fresh (defined blow).

\vspace{0.2cm}

At the end of the game, the adversary $\mathcal{A}$ outputs a bit $b'$ as her guess for $b$. The advantage of $\mathcal{A}$ winning the game is defined as $Adv_{\mathcal{A}}^{clke}(k)=2Pr[b'=b]-1$.

As an instance $\prod_{U}^{i}$ uses both  long-term key pair $((ID_{U},pk_{U}),sk_{U})$ and ephemeral key pair $(epk_{U},esk_{U})$, once both the $sk_{U}$ and $esk_{U}$ are exposed, the adversary can trivially compute the session key $ssk_{U}^{i}$. The instance $\prod_{U}^{i}$ is safe if none of the conditions is true:

\vspace{0.1cm}

(1) The adversary makes a RevealSessionKey$(U,i)$ query.

\vspace{0.1cm}

(2) The adversary makes both RevealSecretKey$(U)$ and RevealEphemeralKey$(U,i)$ queries.

\vspace{0.1cm}

(3) The adversary makes RevealMasterKey query or RevealIDBasedKey$(U)$ query, and also makes both RevealSecretValue$(U)$ query and RevealEphemeralKey$(U,i)$ queries.

\vspace{0.1cm}

(4) $\prod_{U}^{i}$ uses a $public/private$ key pair which is different from its original key pair, and the adversary makes RevealMasterKey query or RevealIDBasedKey$(U)$ query, and also makes a RevealEphemeralKey$(U,i)$ query.

\vspace{0.2cm}

\noindent$\displaystyle {\bf Definition 1.} $  Session  Freshness

\vspace{0.1cm}

Let $\prod_{U}^{i}$ denote an instance with $acc_{U}^{i}=ture$ and $pid_{U}^{i}=V$. If any of the following conditions is true, the $\prod_{U}^{i}$ is unfresh.

\vspace{0.1cm}

(1) $\prod_{U}^{i}$ is exposed.

\vspace{0.1cm}

(2) $\prod_{U}^{i}$ has a partner instance $\prod_{v}^{j}$, and $\prod_{v}^{j}$ is exposed.

\vspace{0.1cm}

(3) If the $\prod_{U}^{i}$ has no partner instance, and either of the following cases happens:

\vspace{0.1cm}

\ \ (a) the adversary makes RevealMasterKey query or RevealIDBasedKey$(V)$ query, and

\ \ \ \ \ makes a RevealSecretValue$(V)$ query;

\vspace{0.1cm}

\ \  (b) the adversary makes a RevealSecretKey $(V)$ query;

\vspace{0.1cm}

\ \ (c) the adversary makes RevealMasterKey query or RevealIDBasedKey$(V)$ query, and

\ \ \ \  makes a ReplacePK$(V,U,i)$ $query\backslash request$.

\vspace{0.2cm}

\noindent$\displaystyle {\bf Definition 2.} $  A CL-KE protocol is said to be secure if

\vspace{0.1cm}

(1) in the presence of a benign adversary who only faithfully conveys messages, then two instances output the same session key;

\vspace{0.1cm}

(2) for any PPT adversary, $Adv_{\mathcal{A}}^{clke}(k)$ is negligible.

\vspace{0.2cm}

\noindent$\displaystyle {\bf Definition 3.} $  Forward Secrecy

\vspace{0.1cm}

Forward secrecy means that learning the full user secret key should not allow an adversary to derive an already-established session key.

\vspace{0.3cm}

\section{Review of Yang and Tan's CL-KE proocol}
\label{3}

\vspace{0.1cm}

Yang and Tan's certificateless key exchange protocol without pairing [20] consists of six algorithms: {\bf Setup}, {\bf ExtractIdBasedKey}, {\bf SetSecretValue}, {\bf SetPublicKey}, {\bf SetPrivateKey} and {\bf Key Exchange}. which is described as follows:

\vspace{0.1cm}

Let $DS= \{ KG,Sig,Ver \}$ denote a digital signature scheme that is unforgeable under adaptive chosen-message attack [21].

\vspace{0.1cm}

{\bf Setup $(1^{k})$ }. KGC chooses a cyclic group $G$ of prime order $q$, and picks a random number $x \in Z_{q}$, and $g \in G \setminus \{1\} $, and computes $g^{x}=y$ . Then, KGC runs the key generation algorithm of $DS$ to generate a signature/verfication key pairing $(sk,vk)$. At last, KGC sets $msk=(x,sk),mpk=(y,vk)$.

\vspace{0.2cm}

{\bf ExtractIdBasedKey$(msk,ID)$ }. Given an identity $ID$, KGC picks a random number $a\in Z_{q}$, computes $R_{ID}=g^{a},z_{ID}=a+H_{1}(ID||R_{ID})x\ mod\ q$, generates a signature $\delta_{ID}=Sig(sk,ID||R_{ID})$ and sets $D_{ID}=(R_{ID},\delta_{ID},z_{ID})$.

\vspace{0.2cm}

{\bf SetSecretValue$(mpk,ID)$ }. The user with identity $ID$ randomly selects $t\in Z_{q}$, and sets $S_{ID}=t$.

\vspace{0.2cm}

{\bf SetPublicKey$(mpk,D_{ID},S_{ID})$ }. Given the user's secret value $S_{ID})$, and ID-Based Key $D_{ID}$, the user computes $U_{ID}=g^{S_{ID}}$, and sets $pk_{ID}=(U_{ID},R_{ID},\delta _{ID})$.

\vspace{0.2cm}

{\bf SetPrivateKey$(mpk,D_{ID},S_{ID})$ }. Given the  user's public key $mpk$, secret value $S_{ID})$ and ID-Based Key $D_{ID}$, the user sets $sk_{ID}=(D_{ID},S_{ID})$.

\vspace{0.2cm}

{\bf Key Exchange }. To establish a session key, party $A$ and party $B$ exchange the following messages.

\vspace{0.1cm}

$$A\rightarrow B: ID_{A},pk_{A},E_{A}=g^{e_{A}};$$
$$B\rightarrow A: ID_{B},pk_{B},E_{B}=g^{e_{B}},$$

\vspace{0.1cm}

where $e_{A}\in Z_{q},e_{B}\in Z_{q}$ are randomly selected by $A$ and $B$ respectively.

The computation of the session key between $A$ and $B$ is as follows:

Party $A$: compute

\vspace{0.1cm}

$Z_{1}=E_{B}^{e_{A}}$, \ \ \ $Z_{2}=U_{B}^{S_{A}}$,\ \ \ $Z_{3}=(R_{B}mpk^{H_{1}(ID_{B}||R_{B})})^{z_{A}}$,\ \ \ $Z_{4}=U_{B}^{e_{A}}$,

\vspace{0.1cm}

$Z_{5}=E_{B}^{S_{A}}$,\ \ \  $Z_{6}=(E_{B}R_{B}mpk^{H_{1}(ID_{B}||R_{B})})^{e_{A}+z_{A}}$,\ \ \ \

\vspace{0.1cm}

$Z_{7}=(U_{B}R_{B}mpk^{H_{1}(ID_{B}||R_{B})})^{S_{A}+z_{A}}$.

\vspace{0.1cm}

and output the session key as

\vspace{0.1cm}

$$ssk=H_{2}(sid,Z_{1},Z_{2},Z_{3},Z_{4},Z_{5},Z_{6},Z_{7}),$$

\vspace{0.1cm}

where $sid=ID_{A},ID_{B},pk_{A},E_{A},pk_{B},E_{B}$.

\vspace{0.2cm}

Party $B$: compute

\vspace{0.1cm}

$Z_{1}=E_{A}^{e_{B}}$, \ \ \ $Z_{2}=U_{A}^{S_{B}}$,\ \ \ $Z_{3}=(R_{A}mpk^{H_{1}(ID_{A}||R_{A})})^{z_{B}}$,\ \ \ $Z_{4}=U_{A}^{e_{B}}$,

\vspace{0.1cm}

$Z_{5}=E_{A}^{S_{B}}$,\ \ \  $Z_{6}=(E_{A}R_{A}mpk^{H_{1}(ID_{A}||R_{A})})^{e_{B}+z_{B}}$,\ \ \ \

\vspace{0.1cm}

$Z_{7}=(U_{A}R_{A}mpk^{H_{1}(ID_{A}||R_{A})})^{S_{B}+z_{B}}$.

\vspace{0.1cm}

and output the session key as

\vspace{0.1cm}

$$ssk=H_{2}(sid,Z_{1},Z_{2},Z_{3},Z_{4},Z_{5},Z_{6},Z_{7}),$$

\vspace{0.1cm}

where $sid=ID_{A},ID_{B},pk_{A},E_{A},pk_{B},E_{B}$.

\vspace{0.3cm}

\section{Analysis and improvement of Yang and Tan's protocol}
\label{4}

Yang and Tan [20] claimed that their protocol is provably secure in the random oracle model, including the forward secrecy. That is, if an attacker does not know all of $(D_{A},S_{A},esk_{A})$, or all of $(D_{B},S_{B},esk_{B})$, it is unable for the attacker to derive the session key. However, in this section, we disprove their result by giving concrete attacks, and propose an improved scheme to prevent these attacks.

\vspace{0.2cm}

\subsection{Attack}
\label{4.1}

\vspace{0.1cm}

For this protocol, to derive a session key, an adversary can first make two RevealExtractIDBasedKey queries to learn $z_{A}$ and $z_{B}$, then make a RevealSecretValue$(B)$ query to learn $S_{B}$ and make a RevealEphemeralKey$(A,i)$ query to learn $e_{A}$. Obviously, the adversary learns neither  $S_{A}$ nor $e_{B}$, which satisfy the requirements and Yang and Tan's security model. However, the adversary can also compute the session key. To attack this protocol, the adversary might perform the following steps.

\vspace{0.1cm}

First, the adversary can  compute as follows

\vspace{0.1cm}

$Z_{1}=E_{B}^{e_{A}}$,

\vspace{0.1cm}

$Z_{3}=(R_{B}mpk^{H_{1}(ID_{B}||R_{B})})^{z_{A}}$,

\vspace{0.1cm}

$Z_{4}=U_{B}^{e_{A}}$,

\vspace{0.1cm}

$Z_{6}=(E_{B}R_{B}mpk^{H_{1}(ID_{B}||R_{B})})^{e_{A}+z_{A}}$.

\vspace{0.1cm}

As the adversary can not make a RevealSecretValue(A) query, he can not obtain the value of $S_{A}$, then should not compute the $Z_{2}$, $Z_{5}$ and $Z_{7}$.

\vspace{0.1cm}

However, as to party $B$, the adversary does not obtain the value of $e_{B}$, but learns $S_{B}$ and $z_{B}$, and can compute

\vspace{0.1cm}

$Z_{2}=U_{A}^{S_{B}}$,

\vspace{0.1cm}

$Z_{5}=E_{A}^{S_{B}}$,

\vspace{0.1cm}

$Z_{7}=(U_{A}R_{A}mpk^{H_{1}(ID_{A}||R_{A})})^{S_{B}+z_{B}}$.

\vspace{0.1cm}

It is easy to see that adversary can really derive the session key as

\vspace{0.1cm}

 $$ssk=H_{2}(sid,Z_{1},Z_{2},Z_{3},Z_{4},Z_{5},Z_{6},Z_{7}),$$

 \vspace{0.1cm}

 where $sid=ID_{A},ID_{B},pk_{A},E_{A},pk_{B},E_{B}$.

 \vspace{0.1cm}

The adversary can successfully calculate the session key associated with the calculation between part $A$ and party $B$, which is completely independent and symmetrical.
 \vspace{0.3cm}

\subsection{Our improved scheme}
\label{4.2}

  \vspace{0.1cm}

 From the analysis in the previous section, we can see that the insecurity of Yang and Tan's protocol is due to the independent of the ephemeral key $e_{A},e_{B}$ and the ID-based key $z_{A},z_{B}$ , the $e_{A},e_{B}$ and $z_{A},z_{B}$ are not fully intertwined enough. In the following, we do a slight modification on Yang and Tan's protocol, and show a new CL-KE protocol without pairing which can resist the attack mentioned above. Our improvement is as follows.

{\bf Setup}, {\bf ExtractIdBasedKey}, {\bf SetSecretValue}, {\bf SetPublicKey} and {\bf SetPrivateKey} are the same as those in section 3.

 \vspace{0.1cm}

\noindent$\displaystyle {\bf Key\ \ Exchange} $

 \vspace{0.1cm}

To establish a session key, party $A$ and party $B$ exchange the following messages.

\vspace{0.1cm}

$$A\rightarrow B: ID_{A},pk_{A},E_{A}=g^{e_{A}};$$
$$B\rightarrow A: ID_{B},pk_{B},E_{B}=g^{e_{B}},$$

\vspace{0.1cm}

where $e_{A}\in Z_{q},e_{B}\in Z_{q}$ are randomly selected by $A$ and $B$ respectively.

The computation of the session key between $A$ and $B$ is as follows:

Party $A$: compute

\vspace{0.1cm}

$Z_{1}=E_{B}^{e_{A}}$, \ \ \ $Z_{2}=U_{B}^{S_{A}}$,\ \ \ $Z_{3}=(R_{B}mpk^{H_{1}(ID_{B}||R_{B})})^{z_{A}}$,\ \ \ $Z_{4}=U_{B}^{e_{A}}$,

\vspace{0.1cm}

$Z_{5}=E_{B}^{S_{A}}$,\ \ \  $Z_{6}=(E_{B}R_{B}mpk^{H_{1}(ID_{B}||R_{B})})^{e_{A}+z_{A}}$,\ \ \ \

\vspace{0.1cm}

$Z_{7}=(U_{B}R_{B}mpk^{H_{1}(ID_{B}||R_{B})})^{S_{A}+z_{A}}$,

\vspace{0.1cm}

$Z_{8}=(E_{B}R_{B}mpk^{H_{1}(ID_{B}||R_{B})})^{S_{A}+z_{A}+e_{A}}$,\ $Z_{9}=(U_{B}E_{B})^{S_{A}+e_{A}}$.

and output the session key as

\vspace{0.1cm}

$$ssk=H_{2}(sid,Z_{1},Z_{2},Z_{3},Z_{4},Z_{5},Z_{6},Z_{7},Z_{8},Z_{9}),$$

\vspace{0.1cm}

where $sid=ID_{A},ID_{B},pk_{A},E_{A},pk_{B},E_{B}$.

\vspace{0.2cm}

Party $B$: compute

\vspace{0.1cm}

$Z_{1}=E_{A}^{e_{B}}$, \ \ \ $Z_{2}=U_{A}^{S_{B}}$,\ \ \ $Z_{3}=(R_{A}mpk^{H_{1}(ID_{A}||R_{A})})^{z_{B}}$,\ \ \ $Z_{4}=U_{A}^{e_{B}}$,

\vspace{0.1cm}

$Z_{5}=E_{A}^{S_{B}}$,\ \ \  $Z_{6}=(E_{A}R_{A}mpk^{H_{1}(ID_{A}||R_{A})})^{e_{B}+z_{B}}$,\ \ \ \

\vspace{0.1cm}

$Z_{7}=(U_{A}R_{A}mpk^{H_{1}(ID_{A}||R_{A})})^{S_{B}+z_{B}},$

\vspace{0.1cm}

$Z_{8}=(E_{A}U_{A}R_{A}mpk^{H_{1}(ID_{A}||R_{A})})^{e_{B}+z_{B}}$,\ $Z_{9}=(U_{A}E_{A})^{S_{B}+e_{B}}$.

and output the session key as

\vspace{0.1cm}

$$ssk=H_{2}(sid,Z_{1},Z_{2},Z_{3},Z_{4},Z_{5},Z_{6},Z_{7},Z_{8},Z_{9}),$$

\vspace{0.1cm}

where $sid=ID_{A},ID_{B},pk_{A},E_{A},pk_{B},E_{B}$.




\section{Security discussion}

\label{5}

In this section, we will analyze the security of the improved protocol, and show that it can work correctly. Since our protocol is derived from Yang and Tan's protocol but made appropriate modification, it can achieve forward secrecy. Through analysis of the protocol, we show that the protocol can withstand some known attacks, for example, public key replacement attack.

 {\bf 1)\  known- key \ secrecy}

Even if the session key is compromised, the adversary does not compromise past or future sessions, as short-term keys are used in generating session keys. Even the two participants  of the protocol remain the same, it also generate different session keys.

{\bf 2)\  Forward \ secrecy}

Even if the long-term private key is compromised, the adversary does not reveal previously established session keys. Even the adversary obtain  the value of $e_{A}$ and $S_{B}$, he can calculate neither of $z_{8}$ and $z_{9}$, that is, he can not compute the session key, so this protocol can achieve the perfect forward secrecy.

{\bf 3)\  PKG \ forward\  secrecy}

The big advantage of the CL-PKC is no-escrow. Even the PKG's master private key is compromised, the adversary (including the PKG) can not reveal previously established session keys. The adversary may generate partial private key, however, in order to compute the established key, the adversary should also obtain both the value of a short-term private key and the full (long-term) private key.

{\bf 4)\  unknown\  key-share \ resilience}

The aim of this attack is to make one participant believe a key which is shared with another participant, and force the two participants to share the same secret. However, the two participants can never share the same key, for they should use the identifier of the intended peer when they compute the session key.

{\bf 5)\  key-compromise\  impersonation}

Key-compromise impersonation has no work in our proposed protocol. Arming with the private key of A, an adversary can impersonate B to A, however, he can not compute the value of $z_{7}$ without knowing the private key of B.

{\bf 6)\  known \ session-specific\  information\  security}

If the short-term private is compromised, it also does not reveal the established key. Specifically, even an adversary obtains the values of $z_{A}$ and $z_{B}$ in any session between A and B, he can not compute $z_{7}$, $z_{8}$.

\section{Conclusion }
\label{6}

 In this paper, by giving a concrete attack, we have indicated that Yang and Tan's CL-KE protocol without pairing is  not secure under their security model. We have also presented an improvement to prevent the attack and given some further security discussions.

\section*{Acknowledgement}

This work is supported by the National Natural Science Foundation of China (Grant Nos. 61170270, 61100203, 60903152, 61003286, 60821001) and the
 Fundamental Research Funds for the Central Universities (Grant Nos. BUPT2011YB01, BUPT2011RC0505, 2011PTB-00-29, 2011RCZJ15).











\end{document}